\documentclass[useAMS,usenatbib]{mn2e}
\usepackage{natbib}

\usepackage{graphicx}
\usepackage{amsmath}
\usepackage{url}

\usepackage[utf8x]{inputenc}
\usepackage[T1]{fontenc}

\usepackage[varg]{txfonts}
\usepackage{microtype}

\usepackage{pgf}
\usepackage{float}
\usepackage{subcaption}

\renewcommand{\vec}{\mathbf}

\graphicspath{{./figures/}}

\title[Photospheric Logarithmic Spirals]{Photospheric Logarithmic Velocity Spirals as MHD Wave Generation Mechanisms}
\author[S. J. Mumford, R. Erd\'elyi]{
S. J. Mumford$^1$, R. Erd\'elyi$^{1,2}$.\\
$^1$Solar Physics \& Space Plasma Research Centre (SP$^{2}$RC), School of Mathematics and Statistics,\\ The University of Sheffield, Hicks Building, Hounsfield Road, Sheffield, S3 7RH U.K.\\
$^2$ Debrecen Heliophysical Observatory, Research Centre for Astronomy and Earth Sciences,\\ Hungarian Academy of Science, 4010 Debrecen, P.O. Box 30, Hungary.}

\begin{document}
\date{Accepted . Received ; in original form }

\pagerange{\pageref{firstpage}--\pageref{lastpage}} \pubyear{2013}

\maketitle

\label{firstpage}
\begin{abstract}
High-resolution observations of the solar photosphere have identified a wide variety of spiralling motions in the solar plasma.
These spirals vary in properties, but are  observed to be abundant at the solar surface.
In this work these spirals are studied for their potential as magnetohydrodynamic (MHD) wave generation mechanisms.
The inter-granular lanes, where these spirals are commonly observed, are also regions where the magnetic field strength is higher than average.
This combination of magnetic field and spiralling plasma is a recipe for the generation of Alfv\'en waves and other MHD waves.

This work employs numerical simulations of a self-similar magnetic flux tube embedded in a realistic, gravitationally stratified, solar atmosphere to study the effects of a single magnetic flux tube perturbed by a logarithmic velocity spiral driver.
The expansion factor of the logarithmic spiral driver is varied and multiple simulations are run for a range of values of the expansion factor centred around observational constraints.

The simulations are analysed using `flux surfaces' constructed from the magnetic field lines so that the vectors perpendicular, parallel and azimuthal to the local magnetic field vector can be calculated.
The results of this analysis show that the Alfv\'en wave is the dominant wave for lower values of the expansion factor, whereas, for the higher values the parallel component is dominant.
This transition occurs within the range of the observational constraints, meaning that spiral drivers, as observed in the solar photosphere, have the potential to generate a variety of MHD wave modes.
\end{abstract}


\section{Introduction}\label{sec:intro}
The solar atmosphere is a highly dynamic, often unpredictable and turbulent environment.
It also has a direct impact on the Earth, with events such as Coronal Mass Ejections causing geomagnetic storms which can disrupt systems such as satellites and power grids.
The outermost layer of the solar atmosphere, the corona, is observed to be heated to millions of degrees Kelvin.
This hot plasma requires a constant energy input to prevent it from cooling, and the mechanism by which this energy is transferred into the corona is a subject of intense study.
The source of the energy is in the photosphere and the internal deeper regions of the Sun.

The solar photosphere is a highly dynamic region of the solar atmosphere, with hot plasma rising up from the convection region, radiating and sinking back down within inter-granular lanes.
Combined with this are multi-scale magnetic fields which intersect the photosphere and one outcome is the generation of magnetohydrodynamic (MHD) waves by various driving motions and at different frequencies.
These MHD waves are generated in a variety of vertical or near vertical magnetic structures, which connect the different layers of the gravitationally stratified solar atmosphere.
This yields a potential mechanism for energy transport vertically through the solar atmosphere, along these magnetic structures, which has been widely studied as a potential solution to the coronal heating problem.
This, MHD wave heating of the solar atmosphere, has been studied analytically \citep[\textit{e.g.}][]{andries2009,wang2011}, observationally \citep[\textit{e.g.}][]{bogdan2006,kobanov2006,morton2012,jess2009,taroyan2009, dorotovic2014} and numerically \citep[\textit{e.g.}][]{bogdan2003,hasan2008,scullion2011,fedun2011,vigeesh2012,Wedemeyer-Bohm2012}

This work, as a follow-up to \cite{Mumford2015}, investigates the effect of logarithmic spiral-type velocity drivers in the solar photosphere and their properties as MHD wave generation mechanisms.
\cite{Mumford2015} studied five different photospheric velocity fields as drivers for MHD waves.
Three of the five drivers considered were spiral type drivers, based on observations of spiral motions in the solar atmosphere \citep{Bonet2008,Wedemeyer-Bohm2009,Bonet2010,Wedemeyer2013}, these motions were modelled as circular, Archemedian and logarithmic motions.
It was concluded that the logarithmic, Archemedian and uniform spiral drivers all generate similar ($\pm 10\%$) excited energy fluxes.
The spiral expansion factors were selected arbitrarily in \cite{Mumford2015}.
This work analyses the effects of the spiral expansion factor on the MHD waves generated by the logarithmic spiral driver, motivated by the observational studies and constraints of \cite{Bonet2008}.
In \cite{Bonet2008} magnetic bright points (MBPs) were observed spiralling in an inter-granular lane, where cold plasma sinks down into the convection zone.
\cite{Bonet2008} fit the observed locations of the MBP with time to the equation for a logarithmic spiral, shown in Equation (\ref{eq:log_spiral}),
\begin{equation}
	\theta = \frac{1}{B_L}\ln(r/a),
	\label{eq:log_spiral}
\end{equation}
where $r$ is the radius of the spiral and $a$ is a positive real constant, and obtained a value of $B_L^{-1} = 6.4 \pm 1.6$ or $B_L = 0.15$ for the dimensionless expansion factor parameter.

In \cite{Bonet2010} a larger sample of photospheric vortices were studied, despite not fitting spirals to the observed motions, a number density of photospheric vortices was calculated as $d \simeq 3.1 \times 10^{-3}$ vortices Mm$^{-2}$ minute$^{-1}$, which therefore provides an upper limit of the number of logarithmic spiral-like vortices in the solar photosphere.

\begin{figure}
	\pgfimage[width=\columnwidth]{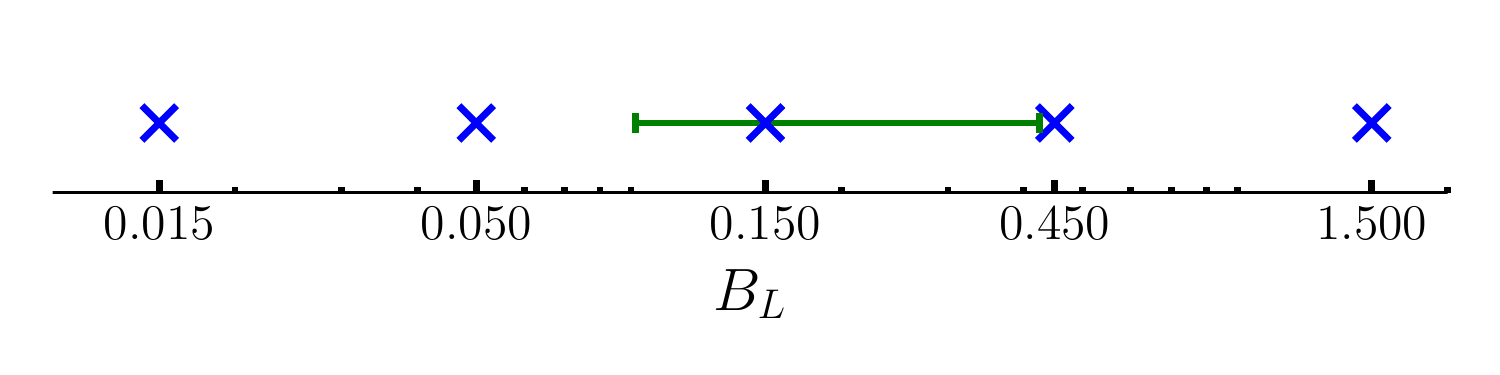}
\caption{The parameter space of $B_L$ used in this work shown as blue crosses, with the $x$-axis on a logarithmic scale. The green error bars show the fit uncertainty of the value observed by \citet{Bonet2008}.}
	\label{fig:B_L_values}
\end{figure}

In this work we investigate the role of the spiral expansion factor ($B_L$) in the generation of MHD waves in a non-potential Gaussian magnetic flux tube, embedded in a realistic stratified solar atmosphere.
The observational result of \cite{Bonet2008} is used as a starting point and values $\pm 3\times$ and $\pm 10\times$ that value are then employed to give five points in the parameter space, centred around their result, which is illustrated in Figure \ref{fig:B_L_values}.

\section{Simulation Configuration}\label{sec:simconfig}
The simulations performed for this study utilise a realistic stratified solar atmosphere constructed by taking the VALIIIc \citep{vernazza1981} hydrodynamical properties and adding a non-potential self-similar magnetic field.
The self-similar magnetic field configuration is derived from the ones employed by \citet{fedun2011} and recently analytically described in \cite{gent2013, gent2014}; based on \citet{Schluter1958, deinzer1965, low1980, Schussler2005}, and identical to the one in \cite{Mumford2015}.
A magnetic field is constructed via this method, then added to the hydrostatic background and then the pressure balance is satisfied using magneto-hydrostatic equilibrium as described by Equation~(\ref{eq:mhs-condition}), \textit{i.e.}
\begin{equation}
	-(\mathbf{B_b}\cdot \nabla)\mathbf{B_b} + \nabla\left(\frac{\mathbf{B_b}^2}{2}\right) + \nabla p = \rho\mathbf{g},
	\label{eq:mhs-condition}
\end{equation}
where $\vec{B_b}$ is the background magnetic field, $\rho$ is the density, and $p$ is the pressure.
Equation~(\ref{eq:mhs-condition}) corrects the missing negative term in \cite{Mumford2015}, the calculations are not affected as this was a typo.
By using a magnetic footpoint strength of $120$ mT and the background atmosphere as specified by the VALIIIc model, the resulting numerical domain has the plasma $\beta > 1$ at every point.

The Sheffield Advanced Code (SAC) \citep{Shelyag2008} used in this work is configured identically to \cite{Mumford2015}. The domain has a spatial extent of $2.0 \times\ 2.0\ \times\ 1.6$ Mm$^3$ in $x$, $y$ and $z$ respectively, with the origin in the $z$ direction $61$ km above the photosphere. The domain is divided up into $128^3$ grid points giving a physical size of $15.6\ \times\ 15.6\ \times\ 12.5$ km$^3$ for each grid cell.
All of the boundary conditions are open and therefore allow almost all non-linear perturbations to escape without significant reflection.

The magnetohydrostatic background is perturbed during the simulations using a 3D Gaussian weighted logarithmic spiral velocity driver, as described by Equation~(\ref{eq: slog}) \citep{Mumford2015}:
\begin{subequations}
\begin{align}
	V_x &= A \frac{\cos(\theta + \phi)}{\sqrt{x^2 + y^2}}\ e^{-\left(\frac{z^2}{\Delta z^2} + \frac{x^2}{\Delta x^2} + \frac{y^2}{\Delta y^2}\right)} \sin \left(2\pi \frac{t}{P}\right),\\
	V_y &= - A \frac{\sin(\theta + \phi)}{\sqrt{x^2 + y^2}}\ e^{-\left(\frac{z^2}{\Delta z^2} + \frac{x^2}{\Delta x^2} + \frac{y^2}{\Delta y^2}\right)} \sin \left(2\pi \frac{t}{P}\right),\label{eq:Slog}\\
\end{align}
\label{eq: slog}
\end{subequations}
where:
\begin{equation*}
	\theta = tan^{-1}\left(\frac{y}{x}\right),\ \phi = tan^{-1}\left(\frac{1}{B_L}\right),\notag	
\end{equation*}
$A=\frac{20}{\sqrt{3}}$, $\Delta x = \Delta y = 0.1$ Mm and $\Delta z = 0.05$ and $P=180$ s.
Here, $B_L$ is the logarithmic spiral expansion factor discussed in Section~\ref{sec:intro}.

\begin{figure*}
	\centering
	\begin{subfigure}[b]{0.3\textwidth}
		\pgfimage[width=\columnwidth]{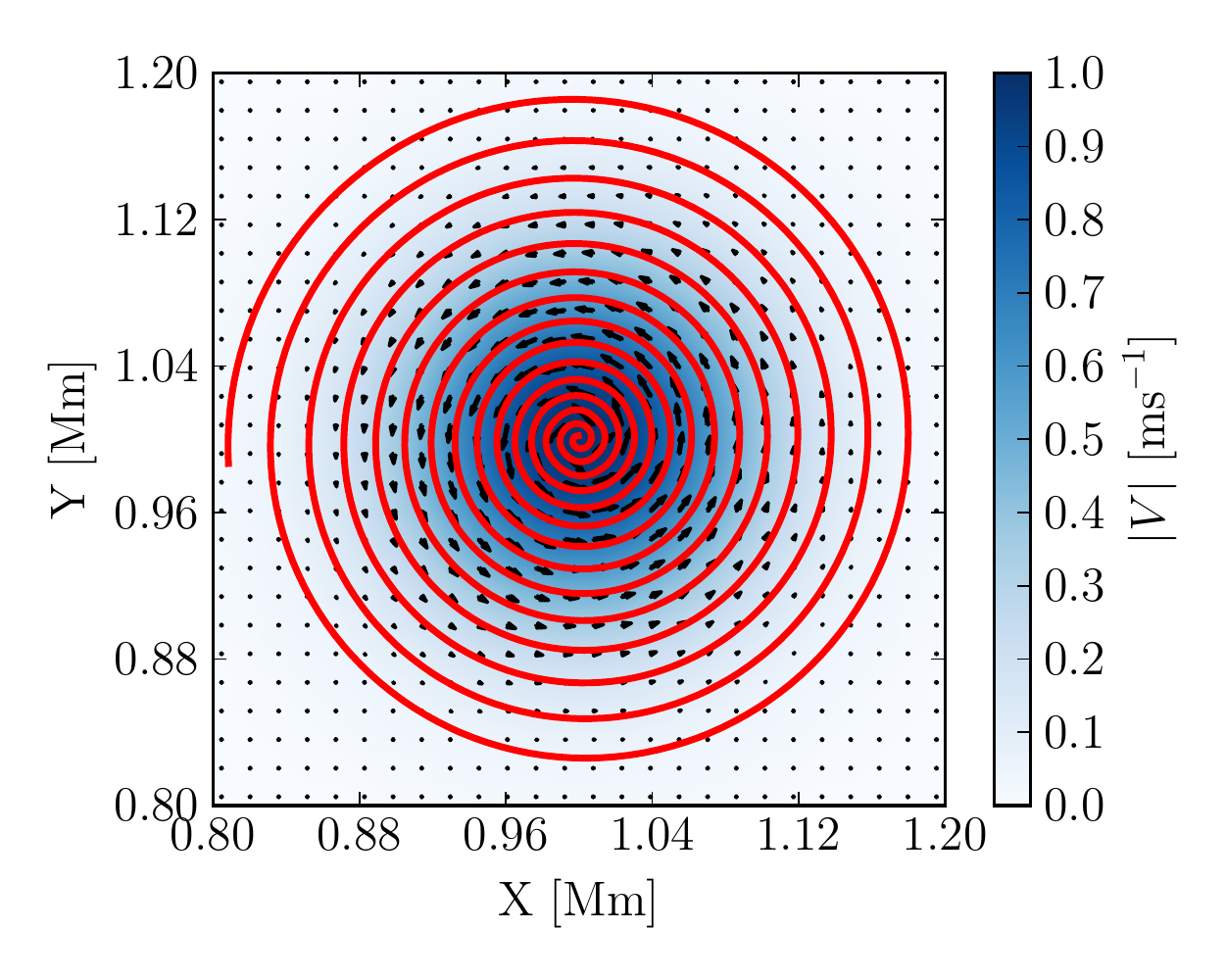}
\caption{$B_L = 0.015$}
\label{fig:All_TD_wave_30:horiz}
	\end{subfigure}
	\begin{subfigure}[b]{0.3\textwidth}
		\pgfimage[width=\columnwidth]{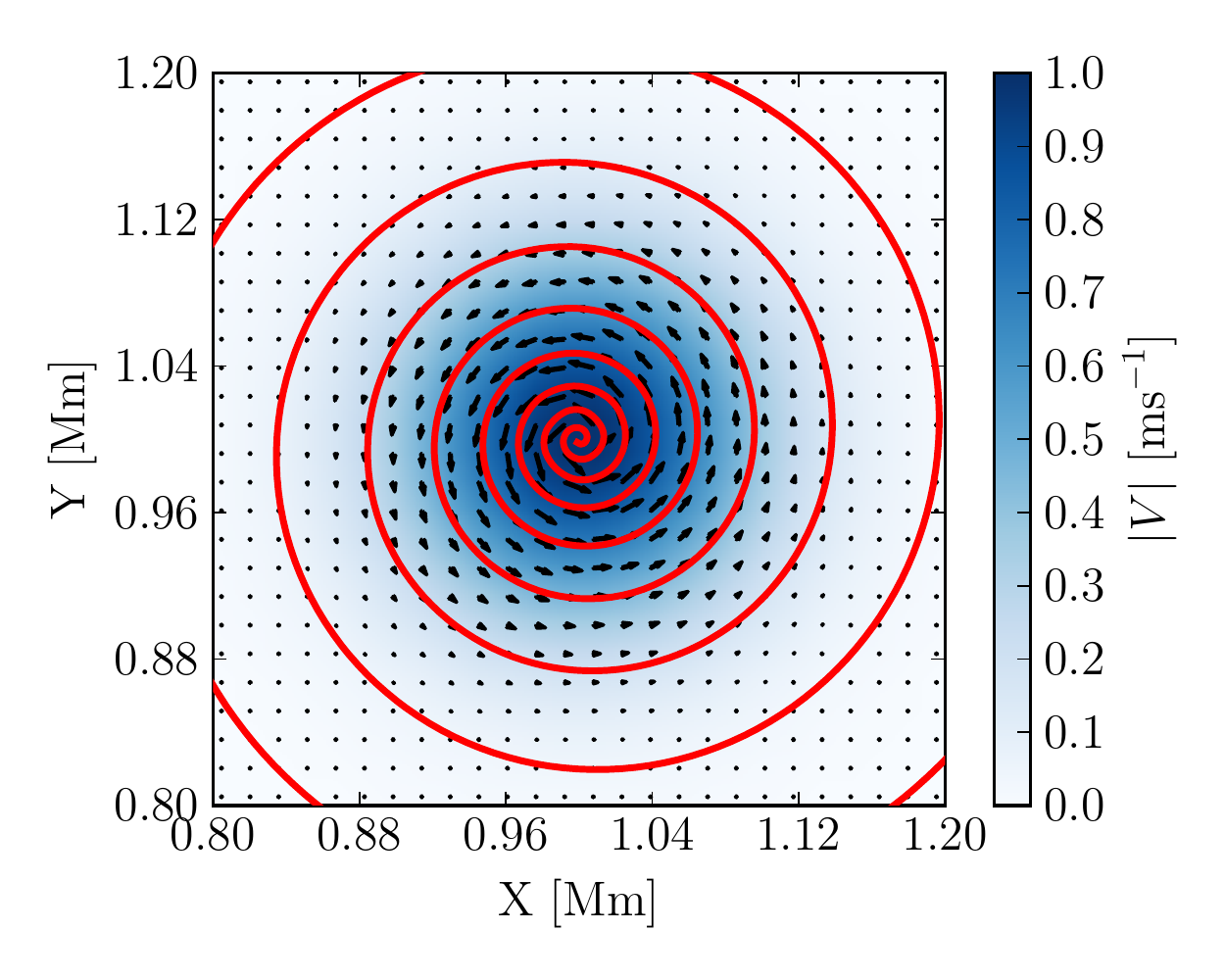}
\caption{$B_L = 0.05$}
\label{fig:All_TD_wave_30:vert}
	\end{subfigure}
	\begin{subfigure}[b]{0.3\textwidth}
		\pgfimage[width=\columnwidth]{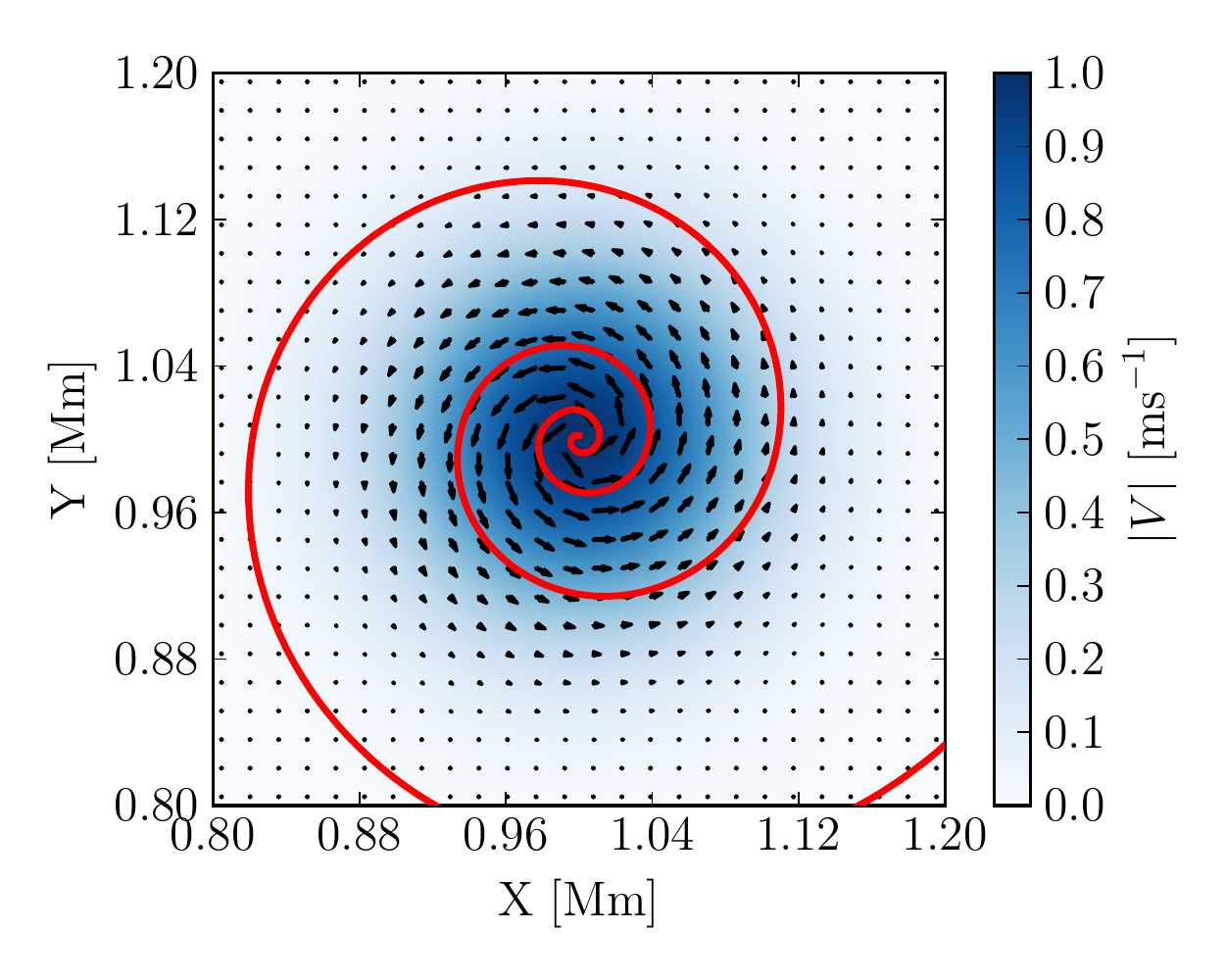}
\caption{$B_L = 0.15$}
\label{fig:All_TD_wave_30:Suni}
	\end{subfigure}
	
	\begin{subfigure}[b]{0.3\textwidth}
		\pgfimage[width=\columnwidth]{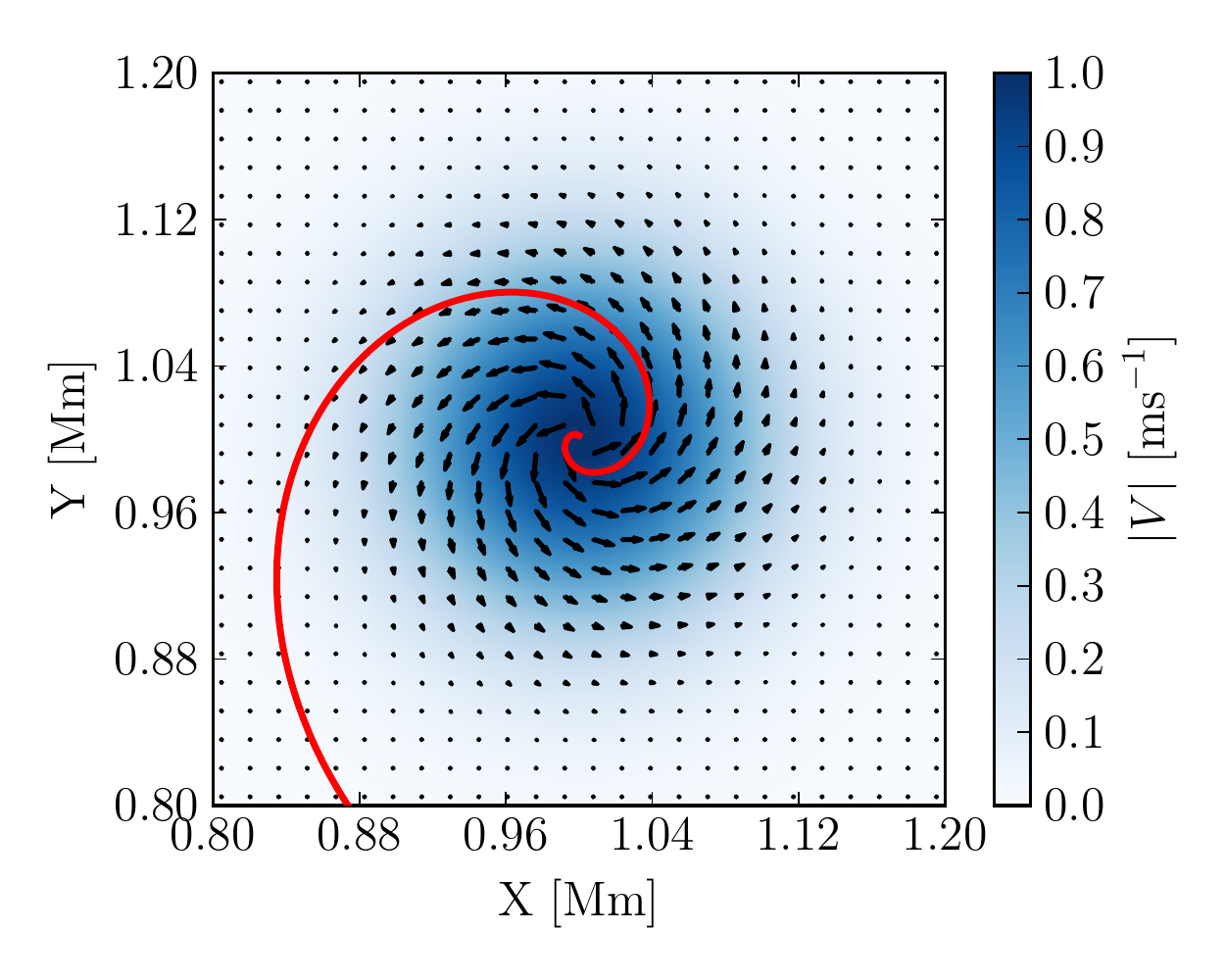}
\caption{$B_L = 0.45$}
\label{fig:All_TD_wave_30:Sarch}
	\end{subfigure}
	\begin{subfigure}[b]{0.3\textwidth}
		\pgfimage[width=\columnwidth]{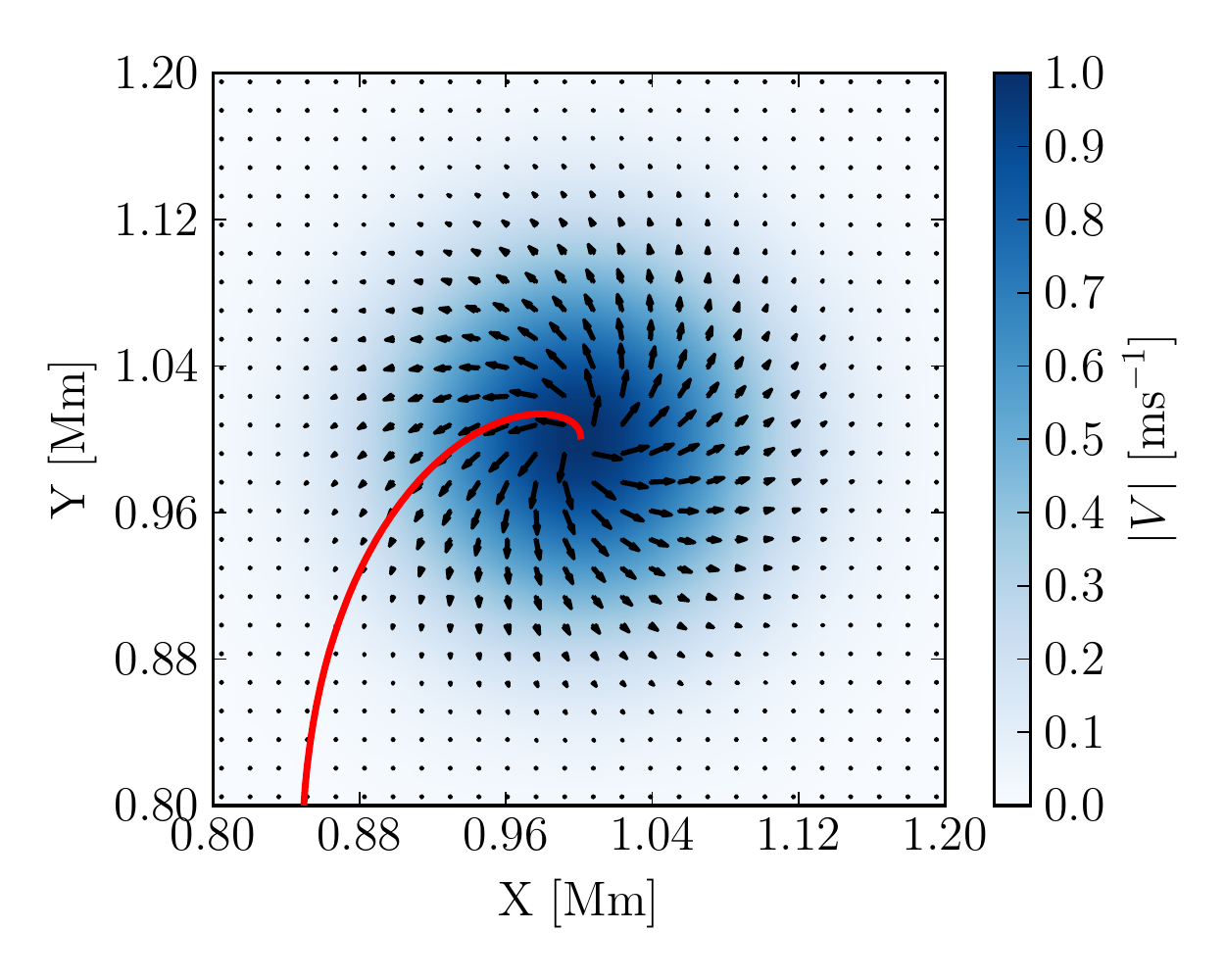}
\caption{$B_L = 1.5$}
\label{fig:All_TD_wave_30:Slog}
	\end{subfigure}
	\caption{Cuts in the [$x$-$y$] plane through the driving velocity field. The normalised velocity is plotted in blue with velocity vectors overplotted in black and a streamline seeded at the centre plotted in red. A plot is shown for each value of $B_L$ used in a simulation.}
	\label{fig:All_log_spirals}
\end{figure*}

Figure \ref{fig:All_log_spirals} shows the calculated velocity profiles for the peak vertical height of the driver ($z=100$ km).
Overplotted on these profiles are streamlines that trace a logarithmic spiral with different expansion factors.

\section{Analysis}\label{sec:analysis}

To quantify the MHD wave modes generated by the logarithmic spiral velocity drivers it is necessary to quantify the relative proportion of the excited MHD wave modes.
The modes present in the domain are assumed to be uniquely determined by the three wave modes present in a uniform homogeneous plasma, namely, the fast magnetoacoustic mode, the slow magnetoacoustic mode and the Alfv\'en mode.
The numerical domain used in this work has the plasma $\beta > 1$ everywhere, therefore, we consider wave propagation in this regime.
Under these conditions the three MHD wave modes are separable into three vector components of perturbation with respect to the magnetic field.
The fast magnetoacoustic mode is the dominant mode in the parallel vector component with respect to the magnetic field.
The slow magnetoacoustic mode is the dominant contributor to the vector component perpendicular to both the magnetic field vector and to the magnetic flux surfaces.
The Alfv\'en mode can be identified in the third vector component, found via the cross product of the parallel and perpendicular vector, parallel to the flux surface but perpendicular to the magnetic field vector.
However, plasma geometry and conditions in the simulation domain make this approximation somewhat imperfect, because there are no clear MHD eigenmodes due to the physical coupling of the waves.
Further, these three modes become degenerate in cylindrical geometry giving rise to sausage, kink, and fluting modes.
Also, due to the complex plasma conditions in the simulation domain the modes may become physically coupled meaning that it is impossible to completely separate the modes.
Despite these complications the description of the modes based on the three vector components in the magnetic field frame is taken as a good way to describe, identify and quantify the MHD wave modes in the system.

To identify theses waves via the vector components relative to the magnetic field the identification of a vector perpendicular to the magnetic field vector is required.
In a 2D system this is a trivial step, however, in a 3D simulation it is ill-defined.
The solution to this problem, used in this work, is to define a magnetic flux surface which encapsulates a constant amount of magnetic flux at all heights in the domain.
This method is described in more detail and utilised in \cite{Mumford2015}.
The surface then allows the computation of a vector perpendicular to it and, thus, to the magnetic field lines it is constructed from.
These `flux surfaces' are initially constructed from a ring of axisymmetric field lines computed in the static background conditions.
The field line seed points then move with the plasma velocity throughout the simulation, which results in the flux surface being constructed from the same field lines at all times in the simulation.
The combination of the surface normal vectors and the magnetic field vector provide the information required to calculate the azimuthal vector via the cross product, which provides a third vector parallel to the surface but perpendicular to the magnetic field.

These surfaces are constructed, using the VTK library\footnote{Visualistation ToolKit 5.10.0 (\url{www.vtk.org})}, for three different characteristic initial radii (measured at the top of the domain) of $156$ km, $468$ km and $936$ km from the centre of the domain, for each simulation, giving a good sampling through the differing plasma properties of the domain.
This allows the analysis of the excited modes at different points in the domain, giving an overall picture of the waves.

Using the flux surfaces, defined above, we can now decompose any vector quantity in the domain into the parallel, perpendicular and azimuthal components, allowing study of the velocity and magnetic field perturbation vectors.
While the velocity and magnetic field perturbation vectors are good for identifying and studying wave behaviour itself, to quantify the amount of each wave mode generated the wave energy flux is computed using Equation (\ref{eq:wave_energy}) from \cite{bogdan2003}.

\begin{equation}
	\vec{F}_{wave} \equiv \widetilde{p}_k \vec{v} + \frac{1}{\mu_0} \left(\vec{B}_b \cdot \vec{\widetilde{B}}\right) \vec{v} - \frac{1}{\mu_0}\left(\vec{v} \cdot \vec{\widetilde{B}} \right) \vec{B}_b,
	\label{eq:wave_energy}
\end{equation}
where subscript $b$ represents a background variable, tilde represents a perturbation from the background conditions and $p_k$ represents kinetic pressure.

The wave energy flux Equation (\ref{eq:wave_energy}) is decomposed onto the flux surface in the same way as the velocity vector, subject to the same limitations as the velocity.

\subsection{Results}\label{subsec:results}

To assist in the visualisation and analysis of the results provided by the flux surfaces the vector components, for both velocity and wave flux, along one field line are extracted for all time steps and plotted as time-distance diagrams in Figures \ref{fig:TD_velocity_r30} and \ref{fig:TD_flux_r30}.

Combining the decomposed velocity vector plotted in Figure \ref{fig:TD_velocity_r30} and the decomposed wave flux vector plotted in Figure \ref{fig:TD_flux_r30} we can reliably describe the nature of the waves generated in the simulations.
Overplotted on all panels in Figures~\ref{fig:TD_velocity_r30} \& \ref{fig:TD_flux_r30} are the phase speeds for the background conditions, the dot-dashed line is the fast speed $v_f$, the dashed line is the sound speed $c_s$, the dotted line is the Alfv\'en speed $v_A$ and the solid line is the slow speed $v_s$.
By comparing these characteristic wave mode speeds to the ridges in the time-distance diagrams it can be seen that in the panels for the torsional component (third panel in each figure), the dominant perturbation travels with the Alfv\'en speed (solid line).
We interpret this perturbation as an Alfv\'en wave.
For the perpendicular component (second panels) it can be seen that the dominant perturbation travels with the slow speed (solid line), therefore this perturbation is interpreted as a slow sausage mode.
We can infer that this perturbation is likely to be a sausage mode perturbation due to the nature of the driver, in that it should not perturb the axis of the flux tube and, that we observe no significant displacement on the flux surfaces during the simulation.
The most interesting result is shown for the parallel component (top panel in each figure), where for lower values of $B_L$, the amplitudes are low, but the perturbations that are present travel with the slow speed (solid line).
However, as $B_L$ increases the perturbations change form.
There seems to appear a second, superimposed perturbation travelling with a speed close to that of the fast (or sound) speeds, which could be a fast sausage mode.
This second perturbation seems to grow proportionally to $B_L$, and can be seen to be dominant in Figures \ref{fig:TD_flux_r30_4} and \ref{fig:TD_flux_r30_5}.

The wave flux graphs in Figure \ref{fig:TD_flux_r30} are components normalised to the magnitude of the wave flux vector, thus showing the relative strengths of the components.
Taking Figure \ref{fig:TD_flux_r30_1} for the $B_L=0.015$ spiral it can be seen that most of the excited wave flux is in the azimuthal component, associated with the Alfv\'en wave.
As the expansion factor ($B_L$) increases, the driver becomes more radial, and the flux starts to shift from the azimuthal component into the parallel component.
This is interpreted as a change of the dominant mode from the torsional Alfv\'en wave into a sausage mode with dominant velocity perturbations parallel to the field lines.
Considering the range of $B_L$, found by \cite{Bonet2008} and illustrated in the range spanned by Figure \ref{fig:TD_velocity_r30_3} and \ref{fig:TD_velocity_r30_4}, it can be seen that even within this parameter range the parallel component becomes substantially more dominant, meaning the change in spectrum of excited MHD wave modes is sensitive to the expansion factor of a spiral driver.

\begin{figure}
	\pgfimage[width=\columnwidth]{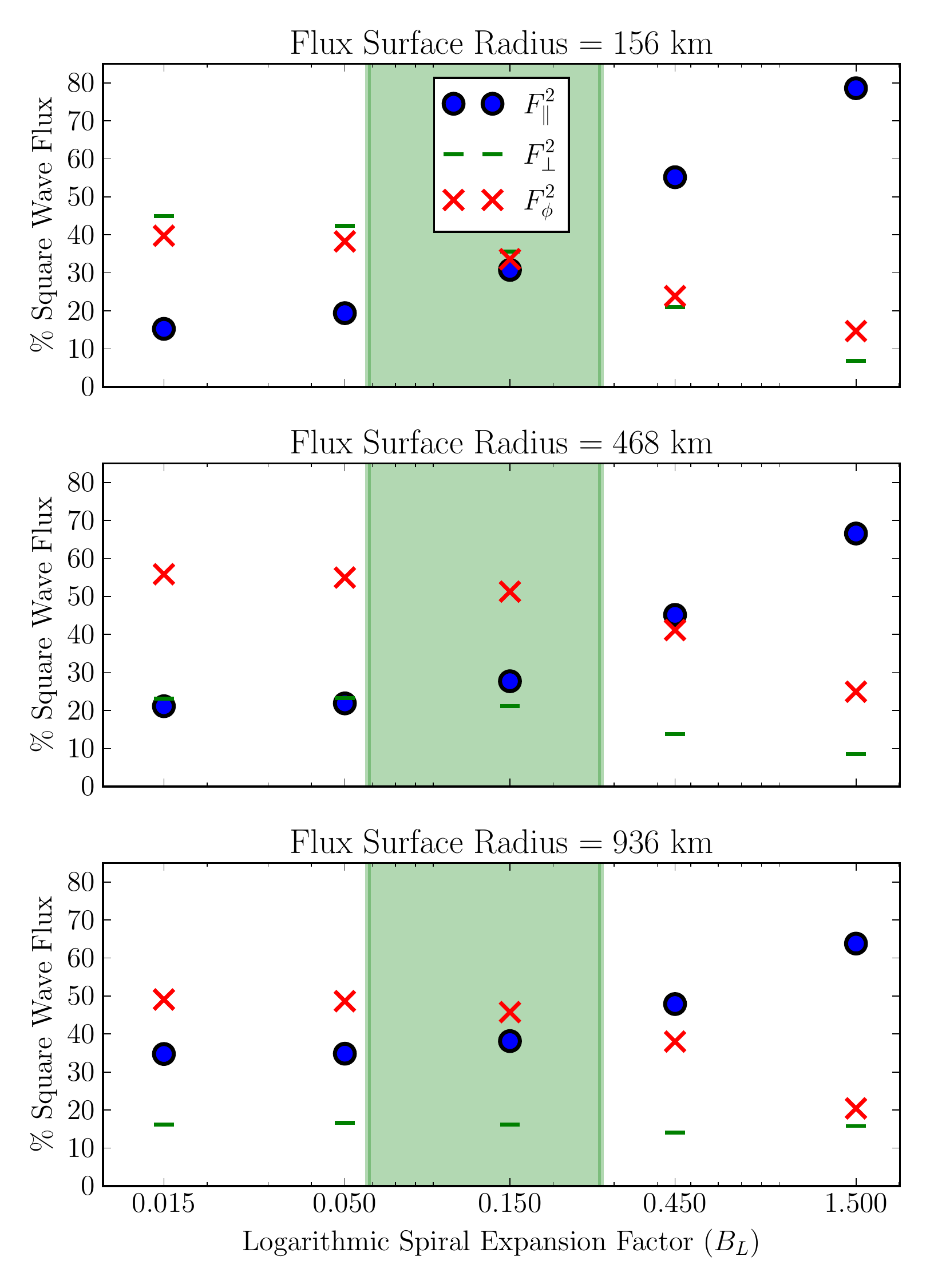}
\caption{Comparison of percentage square wave energy flux for all simulations and all calculated flux surfaces.
	The parallel component of wave energy flux is shown as blue circles, the perpendicular component as green dashes and the azimuthal component as red crosses.
	The green shaded region shows the fit uncertainty in the value observed in \citet{Bonet2008}}
	\label{fig:flux_comparison}
\end{figure}

\cite{Mumford2015} reported that, for the spiral drivers, there is a significant percentage of the wave energy flux contained in the perpendicular component.
This appears to be inversely coupled to the spiral expansion factor of the driver, as it decreases proportionally with the azimuthal wave flux component.
The size of the perpendicular component is also inversely proportional to the initial radius of the flux surface, as can be seen by its decrease in the three panels of Figure \ref{fig:flux_comparison}.

\newcommand{\fwidth}{0.48\textwidth}
\begin{figure*}
	\centering
	
	\begin{subfigure}[b]{\fwidth}
		\pgfimage[width=\columnwidth]{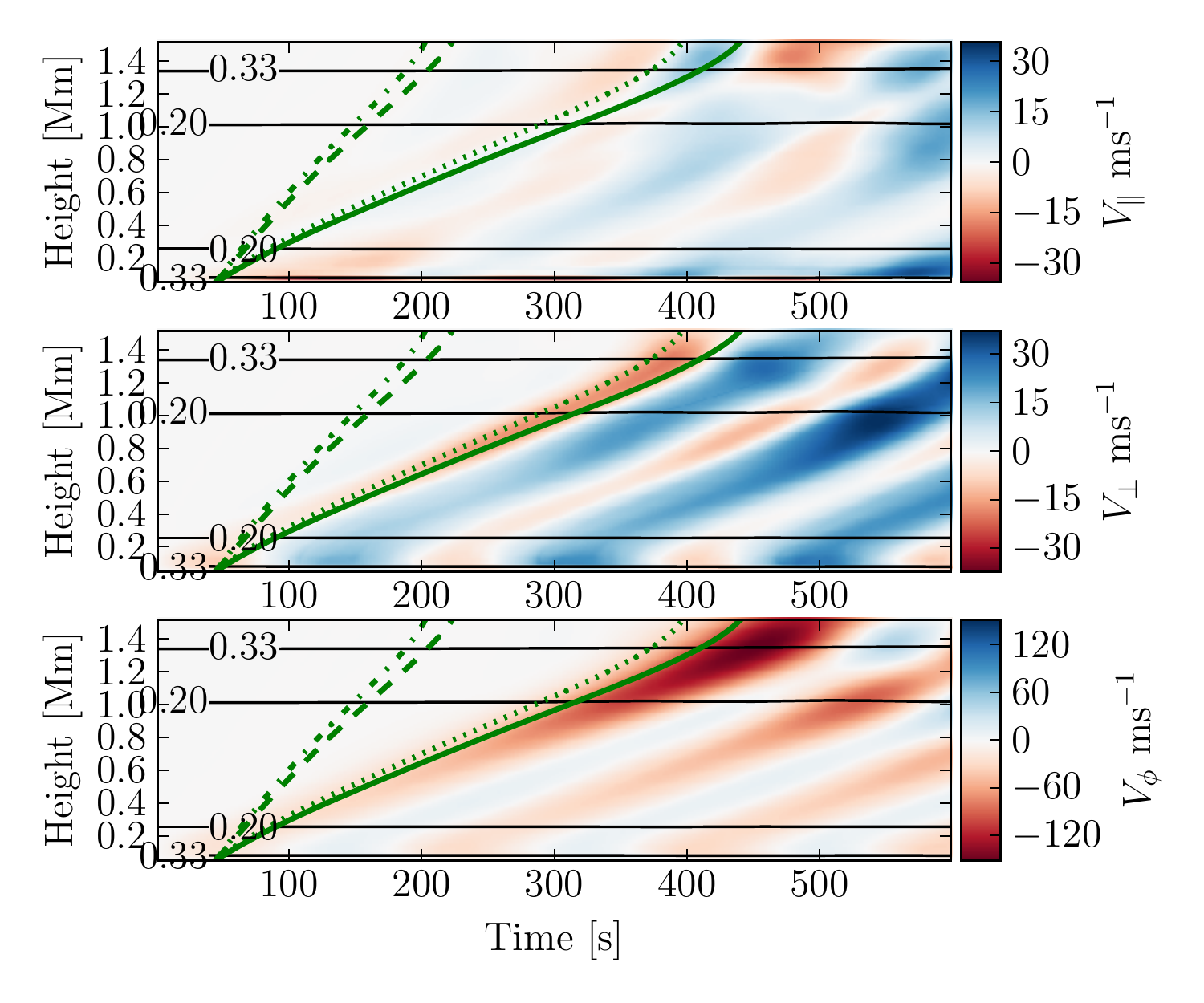}
\caption{$B_L = 0.015$}
\label{fig:TD_velocity_r30_1}
	\end{subfigure}
	\begin{subfigure}[b]{\fwidth}
		\pgfimage[width=\columnwidth]{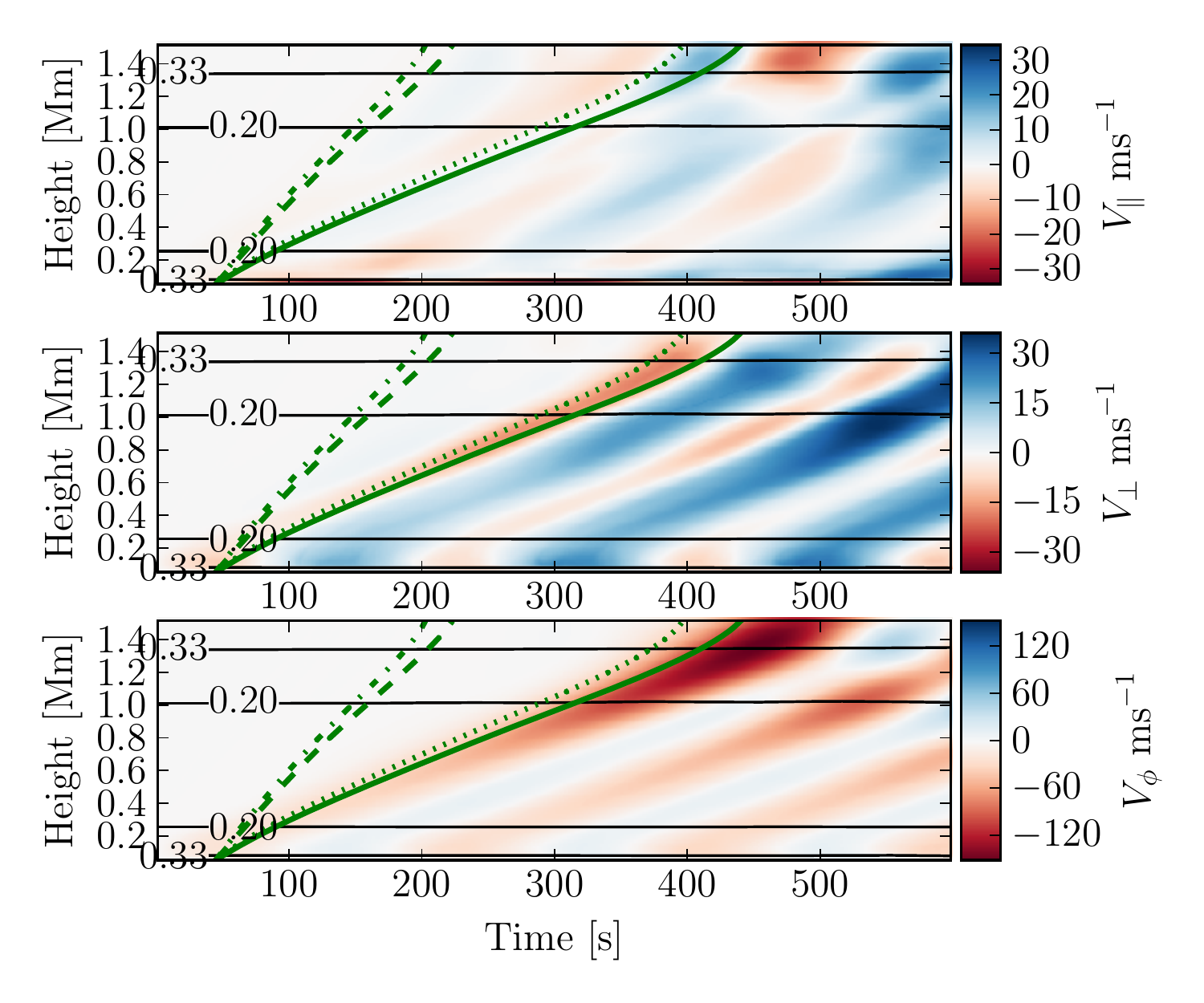}
\caption{$B_L = 0.05$}
\label{fig:TD_velocity_r30_2}
	\end{subfigure}
	
	\begin{subfigure}[b]{\fwidth}
		\pgfimage[width=\columnwidth]{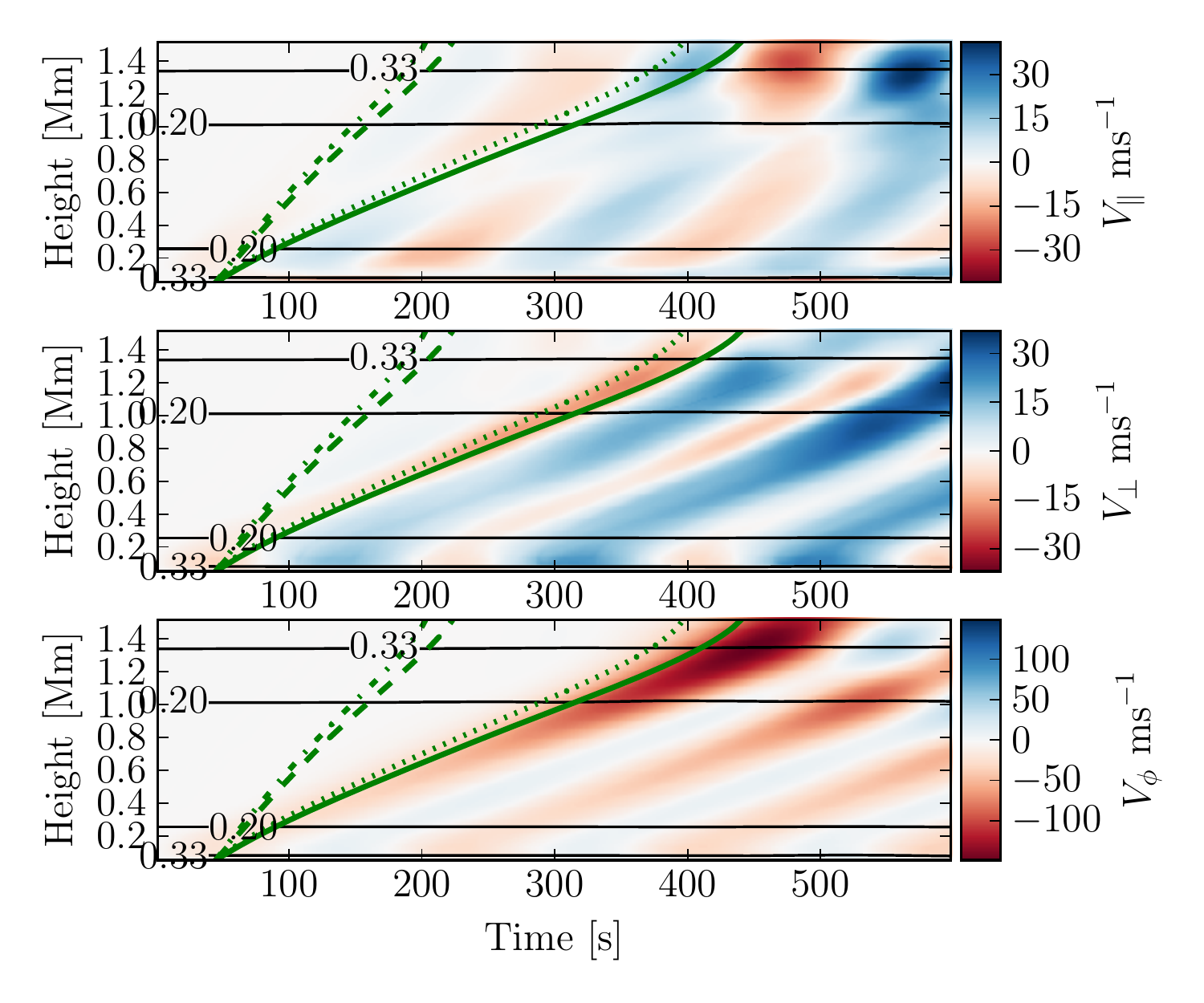}
\caption{$B_L = 0.15$}
\label{fig:TD_velocity_r30_3}
	\end{subfigure}
	\begin{subfigure}[b]{\fwidth}
		\pgfimage[width=\columnwidth]{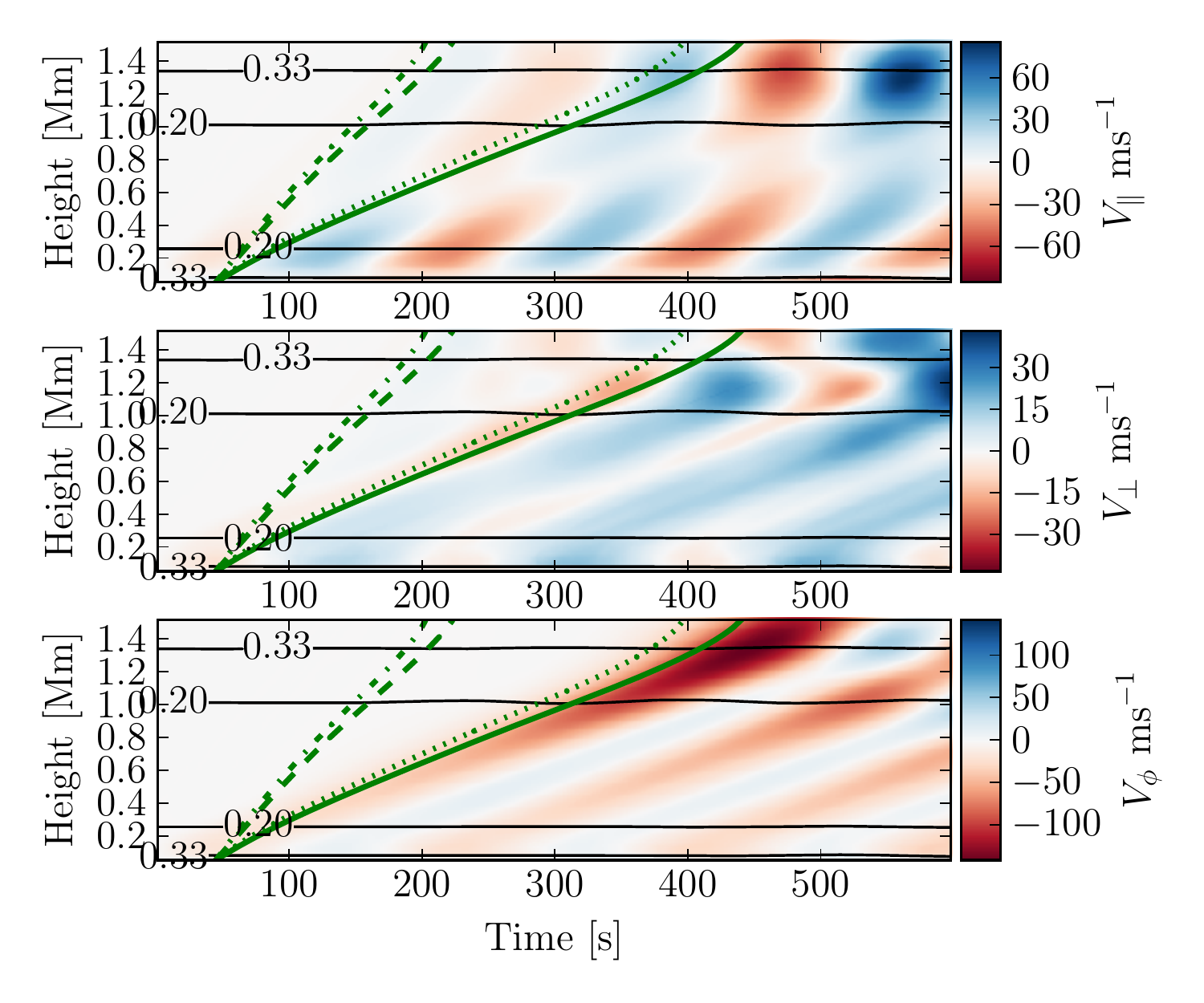}
\caption{$B_L = 0.45$}
\label{fig:TD_velocity_r30_4}
	\end{subfigure}

	\begin{subfigure}[b]{\fwidth}
		\pgfimage[width=\columnwidth]{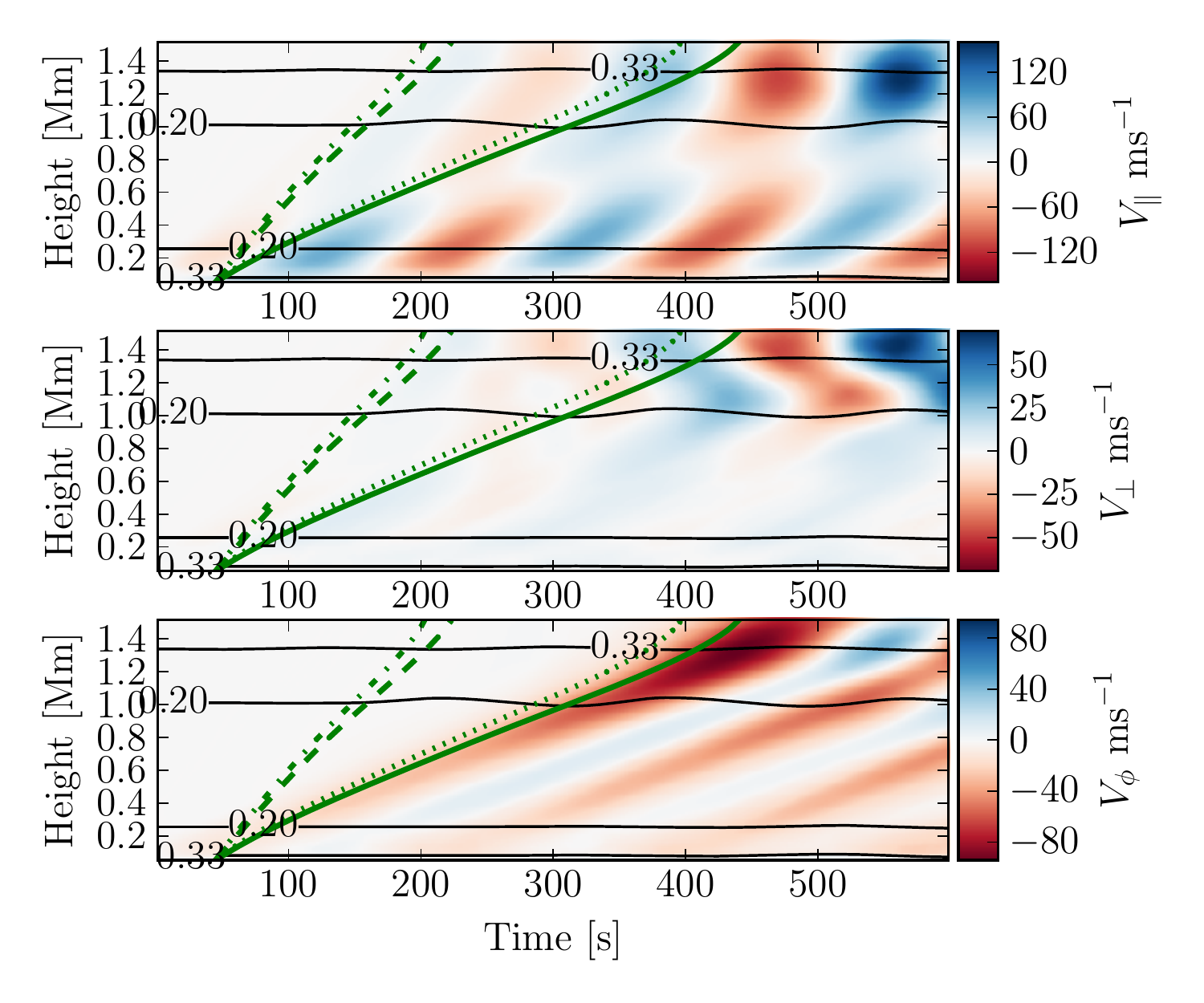}
\caption{$B_L = 1.5$}
\label{fig:TD_velocity_r30_5}
	\end{subfigure}
	\caption{
	Velocity time-distance diagrams for all simulated values of $B_L$ for the surface with an initial top radius of $468$ km.
	Shown in green are the phase speeds for the background conditions, the dot-dashed line is the fast speed $v_f$, the dashed line is the sound speed $c_s$, the dotted line is the Alfv\'en speed $v_A$ and the solid line is the slow speed $v_s$.
	Note that plasma $\beta > 1$ for all heights in the domain.
	}
	\label{fig:TD_velocity_r30}
\end{figure*}

\begin{figure*}
	\centering
	
	\begin{subfigure}[b]{\fwidth}
		\pgfimage[width=\columnwidth]{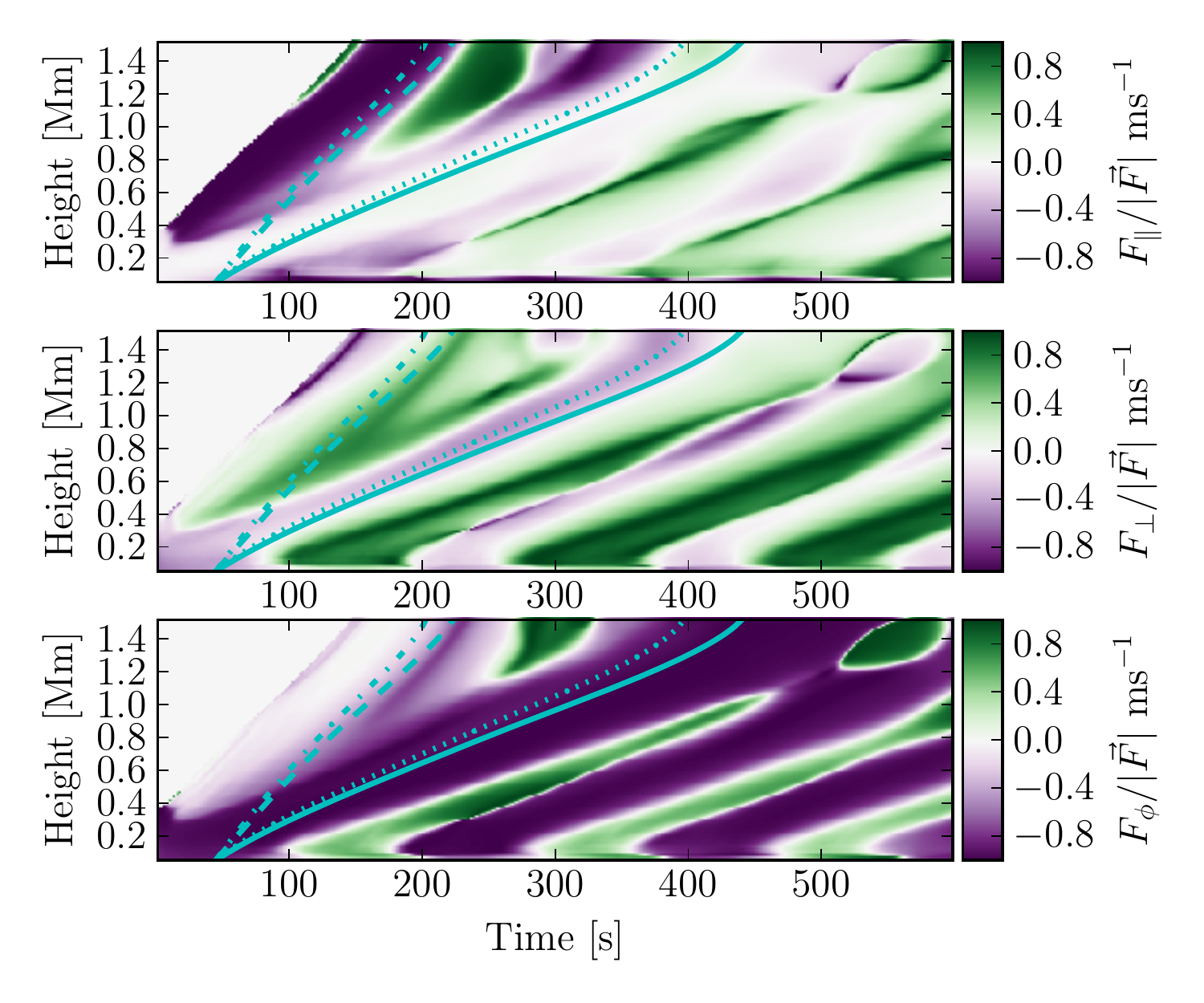}
\caption{$B_L = 0.015$}
\label{fig:TD_flux_r30_1}
	\end{subfigure}
	\begin{subfigure}[b]{\fwidth}
		\pgfimage[width=\columnwidth]{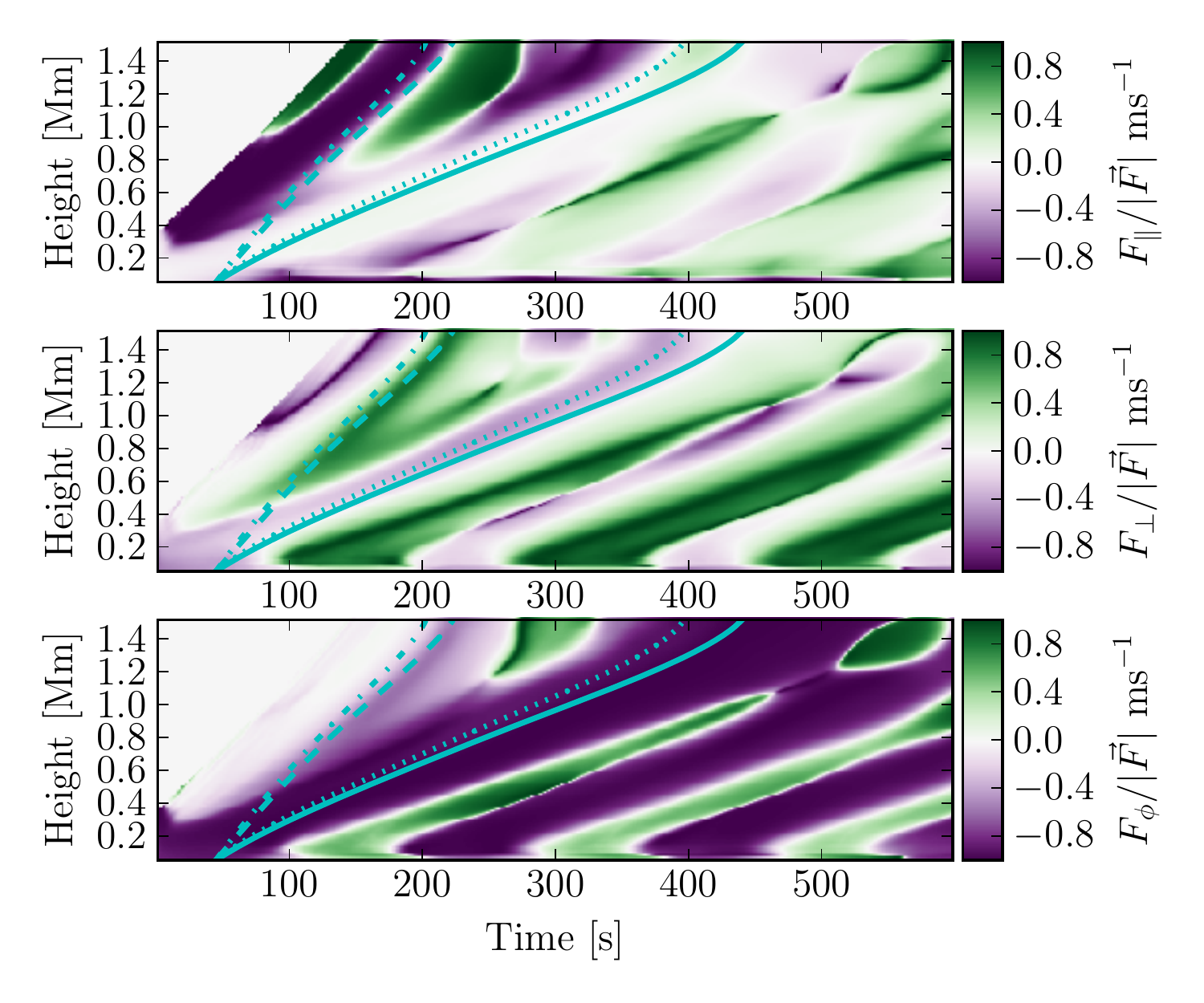}
\caption{$B_L = 0.05$}
\label{fig:TD_flux_r30_2}
	\end{subfigure}
	
	\begin{subfigure}[b]{\fwidth}
		\pgfimage[width=\columnwidth]{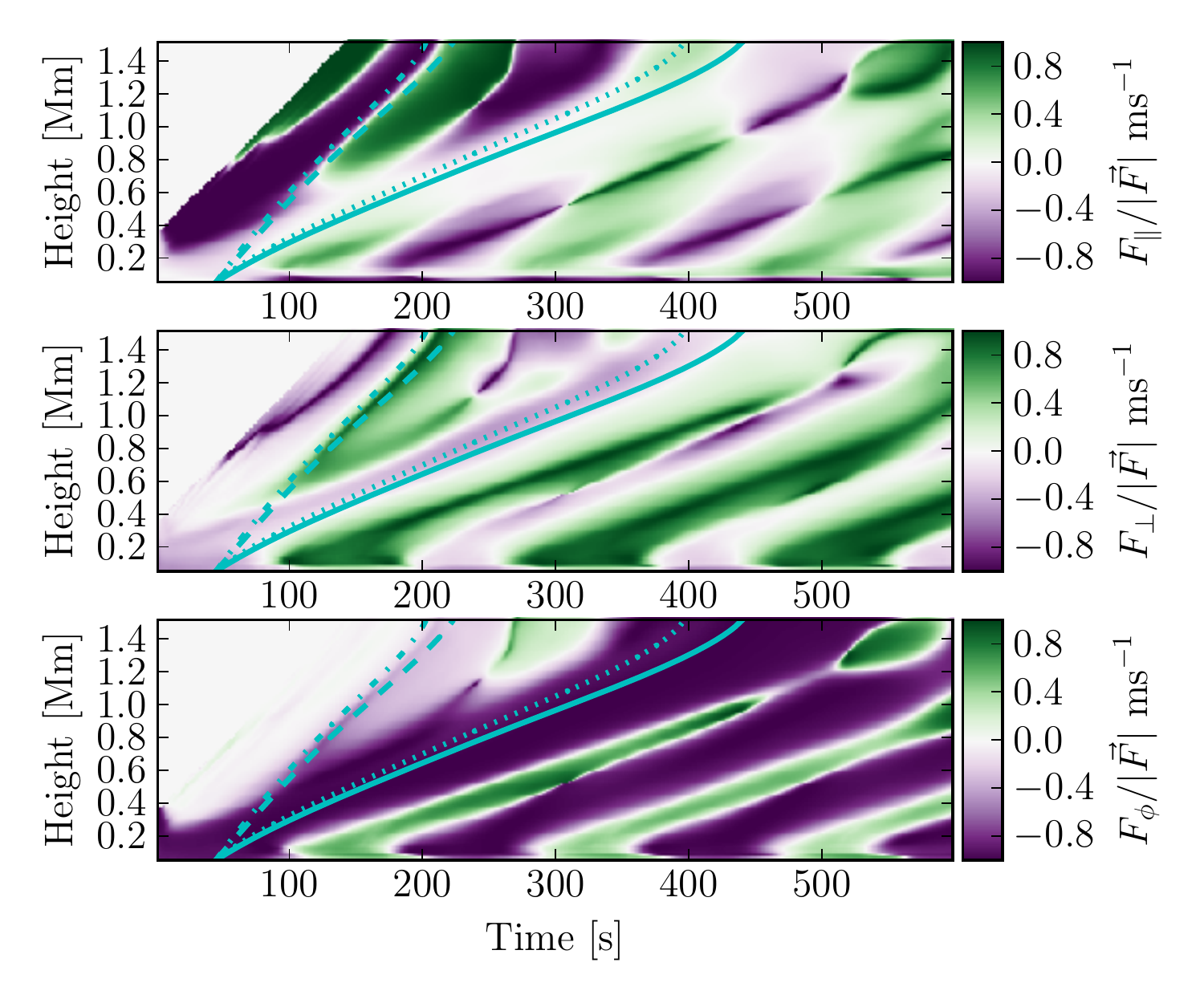}
\caption{$B_L = 0.15$}
\label{fig:TD_flux_r30_3}
	\end{subfigure}
	\begin{subfigure}[b]{\fwidth}
		\pgfimage[width=\columnwidth]{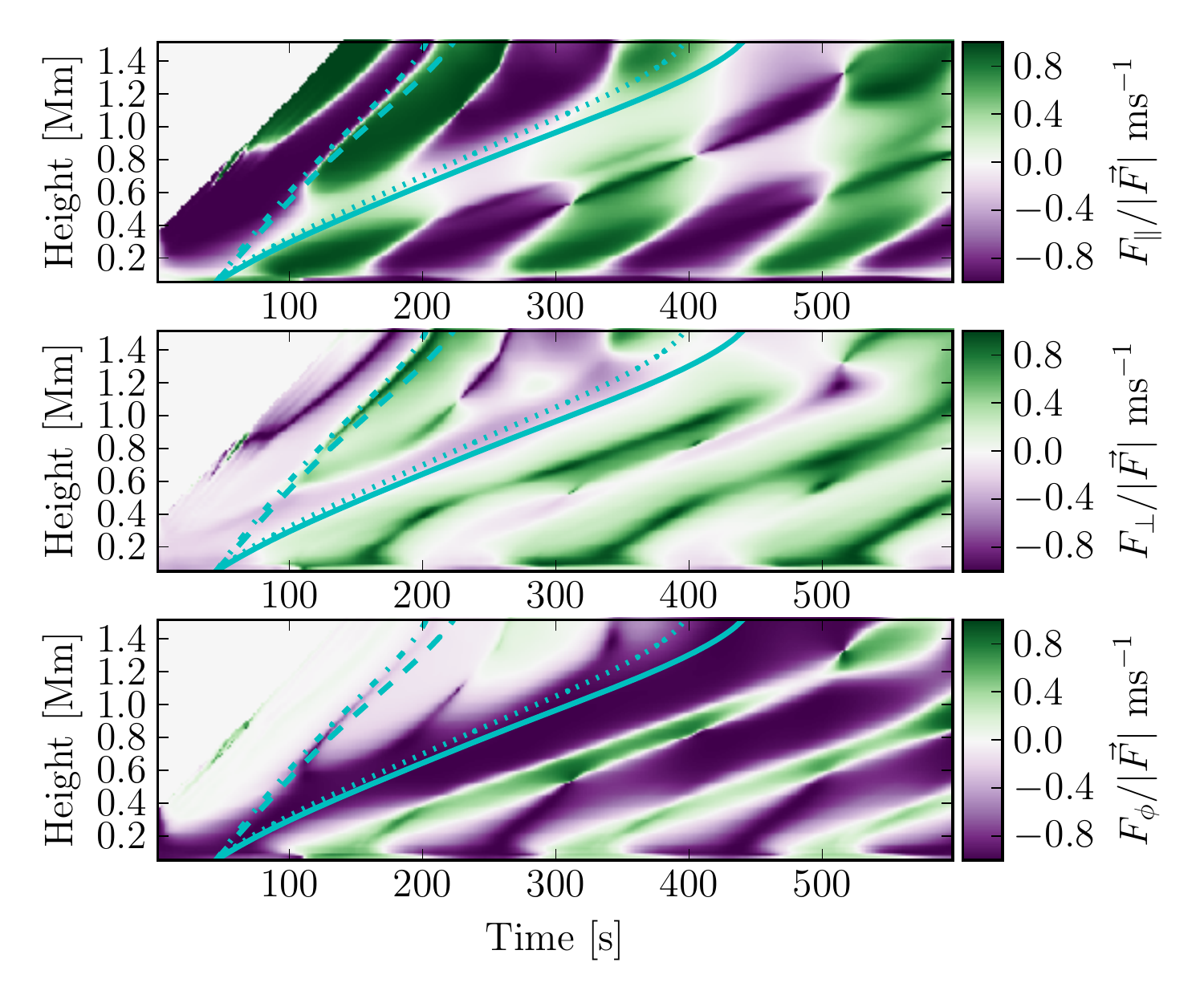}
\caption{$B_L = 0.45$}
\label{fig:TD_flux_r30_4}
	\end{subfigure}

	\begin{subfigure}[b]{\fwidth}
		\pgfimage[width=\columnwidth]{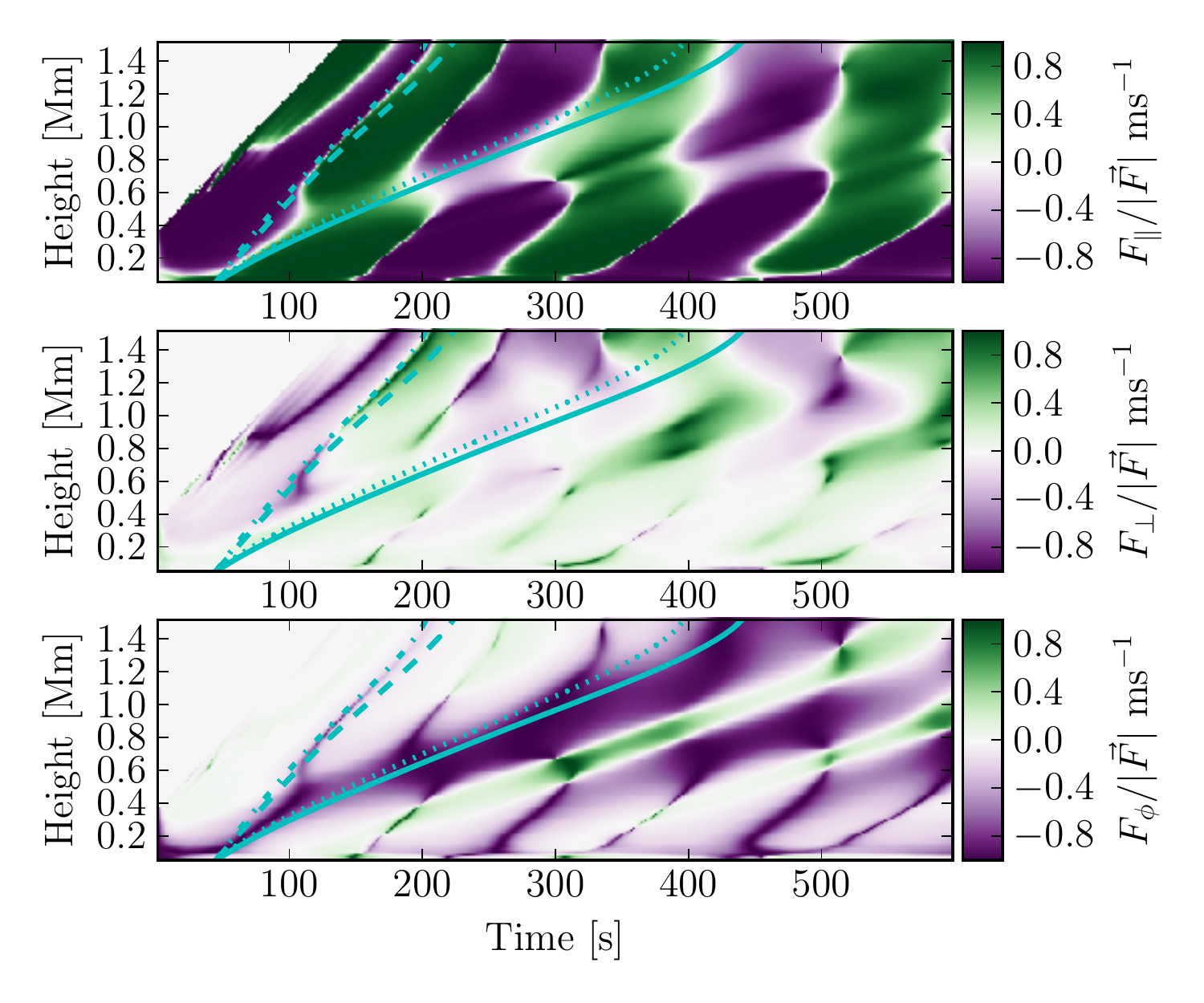}
\caption{$B_L = 1.5$}
\label{fig:TD_flux_r30_5}
	\end{subfigure}
	\caption{
	Normalised wave energy flux time-distance diagrams for all simulated values of $B_L$ for the surface with an initial top radius of $468$ km.
	Shown in blue are the phase speeds for the background conditions, the dot-dashed line is the fast speed $v_f$, the dashed line is the sound speed $c_s$, the dotted line is the Alfv\'en speed $v_A$ and the solid line is the slow speed $v_s$.
	Note that plasma $\beta > 1$ for all heights in the domain.
	}
	\label{fig:TD_flux_r30}
\end{figure*}

This change in excitation of MHD waves is summarised in Figure \ref{fig:flux_comparison}, where the average value of $\displaystyle\frac{F_{\parallel, \perp, \phi}^2}{F_\parallel^2 + F_\perp^2 + F_\phi^2}$ for all time is plotted.
Figure \ref{fig:flux_comparison} shows that, between the values of $B_L=0.15$ and $B_L=0.45$ there is a turning point where the torsional component becomes less dominant, with expansion factors larger than $0.15$ having the parallel component being the dominant component.
This turning point occurs within the range of the fitted spirals in \cite{Bonet2008} and, therefore, implies that photospheric spirals may indeed generate a variety of different MHD modes with varying strengths.

\section{Conclusions}\label{sec:conclusion}
Numerical simulations of the low solar atmosphere were performed where a range of different logarithmic spiral velocity drivers excited MHD waves in a magnetic flux tube.
The expansion factor of the logarithmic spirals was varied around a statistically determined observational value of $B_L^{-1} = 6.4 \pm 1.6$ from \cite{Bonet2008}.
The excited MHD waves were analysed using `flux surfaces' computed from the magnetic field lines, from which physical vector parameters, such as velocity, were decomposed into a reference frame of parallel, perpendicular and azimuthal to the magnetic field lines.
This decomposition enabled the identification of the excited MHD wave modes, and from computing the wave energy flux the relative strengths of the components was calculated.

The average wave flux analysis for all time was used as an indication of the relative strength of the three components, in the magnetic field frame.
The analysis indicates that between $B_L=0.15$ and $B_L=0.45$ the torsional component, associated with the Alfv\'en mode, becomes weaker than the parallel component, which becomes dominant.
Assuming that the expansion rates of these observed spirals will be distributed over a range of possible values, they may not be generating the quantity of Alfv\'en waves and fluxes previously assumed.

\section*{Acknowledgments}
The authors wish to thank V. Fedun for his helpful suggestions on improving the manuscript.
The authors also acknowledge the NumPy, SciPy, Matplotlib \citep{hunter2007}, yt \citep{turk2011} and MayaVi2 \citep{ramachandran2011} Python projects for providing the computational tools to analyse the data. This work made use of the facilities of N8 HPC provided and funded by the N8 consortium and EPSRC (Grant No. EP/K000225/1). The Centre is co-ordinated by the Universities of Leeds and Manchester. RE is thankful to the Science and Technology Facilities Council (STFC) and NSF, Hungary (OTKA, Ref. No. K83133) and acknowledges M. K\'eray for patient encouragement.

\bibliographystyle{mn2e}
\bibliography{smumford_etal_2014}{}
 
\end{document}